\documentclass[3p,preprint]{elsarticle}

\usepackage{amsfonts}
\usepackage{graphicx}

\begin{document}

\begin{frontmatter}

\title{A note on BPS vortex bound states}

\author[aai]{A. Alonso-Izquierdo\corref{cor1}}
\ead{alonsoiz@usal.es}
\author[wgf]{W. Garcia Fuertes}
\ead{wifredo@uniovi.es}
\author[jmg]{J. Mateos Guilarte}
\ead{guilarte@usal.es}

\cortext[cor1]{Corresponding author}

\address[aai]{Departamento de Matematica Aplicada, Universidad de Salamanca, SPAIN}
\address[wgf]{Departamento de Fisica, Universidad de Oviedo, SPAIN}
\address[jmg]{Departamento de Fisica Fundamental, Universidad de Salamanca, SPAIN}

\begin{abstract}
In this note we investigate bound states, where scalar and vector bosons are trapped by BPS vortices in the Abelian Higgs model with a critical ratio of the couplings. A class of internal modes of fluctuation around cylindrically symmetric BPS vortices is characterized mathematically, analysing the spectrum of the second-order fluctuation operator when the Higgs and vector boson masses
are equal. A few of these bound states with low values of quantized magnetic flux are described fully, and their main properties are discussed.
\end{abstract}

\begin{keyword}
Abelian Higgs model \sep Sel-dual (BPS) vortices \sep Boson-Vortex bound states
\end{keyword}

\end{frontmatter}

\section{Introduction}
Very soon after the discovery of Abrikosov quantized flux lines in the Ginzburg-Landau theory of Type II superconductors \cite{Abrikosov}, the existence and nature of fermionic bound states on these vortex filaments were discussed by de Gennes et al. in Reference \cite{deGennes} . Quantized magnetic flux lines were rediscovered by Nielsen and Olesen in the Abelian Higgs model, see \cite{Nielsen}, a finding that enhanced the interest of these topological defects by promoting them to the relativistic and quantum world. By adjusting the couplings in the Abelian Higgs model to drive the system to the critical point between Type II and Type I superconductors, Bogomolny showed, see \cite{Bogomolny}, that quantized vortex lines still exist but move without interaction with respect to each other. Bosonic vortex bound states were investigated by Goodman and Hindmarsh, see \cite{Hindmarsh}, in the context of the Abelian Higgs model for any value of the parameter governing the transition between Type I and Type II superconductivity. In this short note we shall focus on finding BPS vortex bound states and we shall describe these internal boson-vortex modes by a mixture of analytical and numerical methods, at least at the same level of numerical precision as the BPS vortex solutions themselves.

\section{BPS vortex fluctuations}

The Abelian Higgs model describes the minimal coupling between a $U(1)$-gauge field and a charged scalar field in a phase where the gauge symmetry is broken spontaneously . In fact, it is a relativistic version of the Ginzburg-Landau theory of superconductivity. At the transition point between Type I and II superconductivity, where the masses of the Higgs and vector fields are equal, the AHM action reads
\begin{equation}
S[\phi,A]=\int d^4 x \left[ -\frac{1}{4} F_{\mu\nu}F^{\mu \nu} + \frac{1}{2} (D_\mu \phi)^* D^\mu \phi -\frac{1}{8} (\phi^* \phi-1)^2 \right]
\label{action1} \qquad .
\end{equation}
Here, non-dimensional coordinates, couplings and fields are used, while $\phi(x)=\phi_1(x)+i\phi_2(x)$ is a complex scalar field and $A_\mu(x) =(A_0(x),A_1(x),A_2(x), A_3(x))$ is the vector potential. The covariant derivative is defined in the conventional form, $D_\mu \phi(x) = (\partial_\mu -i A_\mu(x))\phi(x)$, whereas the electromagnetic field tensor is also standard: $F_{\mu\nu}(x)=\partial_\mu A_\nu(x) - \partial_\nu A_\mu(x)$. We choose the metric tensor in Minkowski space as $g_{\mu\nu}={\rm diag}(1,-1,-1,-1)$, $\mu,\nu=0,1,2,3$, and use the Einstein repeated index convention. In the simultaneous temporal and axial gauges $A_0=A_3=0$, the Bogomolny arrangement of the energy per unit length for static and $x_3$-independent field configurations $V[\phi,A]$, see \cite{Bogomolny}, shows that solutions of the first order PDE system
\begin{equation}
D_1\phi \pm i D_2 \phi =0 \hspace{0.5cm},\hspace{0.5cm} F_{12}\pm \frac{1}{2} (\phi^*\phi -1)=0 \, \, , \label{ode11}
\end{equation}
with appropriate asymptotic behaviour at infinity in the $\mathbb{R}^2$ $x_1:x_2$-plane, are absolute minima of $V[\phi,A]$. It was proved
in \cite{Taubes} that there exist solutions of the PDE's (\ref{ode11}) with finite string tensions that are proportional to the magnetic flux along the $x_3$-axis of $n\in\mathbb{Z}$ quanta: $V[\phi,A]=\frac{1}{2}\,|\int_{\mathbb{R}^2} d^2 x F_{12}|=\pi | n|$. These topological objects are denoted BPS vortices because they correspond to the Abrikosov-Nielsen-Olesen vortex filaments arising in Type II superconductors, see \cite{Abrikosov}-\cite{Nielsen}, when the scalar and vector penetration lengths in the Ginzburg-Landau free energy are equal and the system lives exactly at the transition point to Type I materials.

Denoting the BPS vortex fields as:
\[
\phi_V=\psi(\vec{x};n)=\psi_1(\vec{x};n) + i \, \psi_2(\vec{x};n) \hspace{0.5cm},\hspace{0.5cm} A_V=(V_1(\vec{x};n),V_2(\vec{x};n)) \hspace{0.5cm}\mbox{with}\hspace{0.5cm} \vec{x}=(x_1,x_2)\quad ,
\]
and assembling the vector and scalar vortex fluctuations $(a_1(\vec{x}),a_2(\vec{x}))$, $\varphi(\vec{x})=\varphi_1(\vec{x})+i\varphi_2(\vec{x})$ in a column vector $\xi(\vec{x})$ with transpose
\[
\xi^t(\vec{x})=\left( \begin{array}{c c c c}a_1(\vec{x}) & a_2(\vec{x}) & \varphi_1(\vec{x}) & \varphi_2(\vec{x}) \end{array} \right)^t \quad ,
\]
one checks that the linearized dynamics is governed by the action of the second-order vortex fluctuation operator ${\cal H}^+$
\begin{equation}
{\cal H}^+= \left( \begin{array}{cccc}
-\Delta + |\psi|^2 & 0 & -2D_1 \psi_2 & 2 D_1 \psi_1 \\
0 & -\Delta +|\psi|^2 & -2 D_2 \psi_2 & 2 D_2 \psi_1 \\
-2 D_1 \psi_2 & -2 D_2\psi_2 & -\Delta +\frac{1}{2} (3|\psi|^2-1)+V_kV_k & -2 V_k \partial_k -\partial_k V_k \\
2D_1\psi_1 & 2 D_2 \psi_1 & 2V_k \partial_k + \partial_k V_k & -\Delta +\frac{1}{2} (3|\psi|^2-1) + V_kV_k
 \end{array} \right)  \label{hmas}
\end{equation}
on $\xi(\vec{x})$. Resolution of the spectral problem ${\cal H}^+ \xi_\lambda(\vec{x}) =\omega_\lambda^2 \, \xi_\lambda(\vec{x})$, where $\lambda$ is a label in either the discrete or the continuous spectrum  of ${\cal H}^+$, permits the decomposition of $\xi(\vec{x})$ as a linear combination of the $\xi_\lambda(\vec{x})$ eigenfunctions.

In the search for bound state (normalizable) eigenfunctions other than zero modes, i.e. $0<\omega_\lambda^2<1$ assuming that $\omega_\lambda^2=1$ is the scattering threshold, we shall profit from a hidden SUSY structure of ${\cal H}^+$. Linear deformation of the PDE system (\ref{ode1}), together with the background gauge
\begin{equation}
\partial_1 [\xi( \vec{x})]_1 + \partial_2 [\xi( \vec{x})]_2-\,\psi_1( \vec{x})\, [\xi( \vec{x})]_4+\psi_2( \vec{x})\,[\xi( \vec{x})]_3 \, \equiv \, \partial_k a_k( \vec{x})-\psi_1( \vec{x})\, \varphi_2( \vec{x})+\psi_2( \vec{x})\,\varphi_1( \vec{x})=0 \quad ,
\label{backgroundgauge}
\end{equation}
is encoded in the following first-order PDE operator
\begin{equation}
{\cal D}= \left( \begin{array}{cccc}
-\partial_2 & \partial_1 & \psi_1 & \psi_2 \\
-\partial_1 & -\partial_2 & -\psi_2 & \psi_1 \\
\psi_1 & -\psi_2 & -\partial_2 + V_1 & -\partial_1 -V_2 \\
\psi_2 & \psi_1 & \partial_1+V_2 & -\partial_2 + V_1
\end{array} \right)  \label{zeromode1} \, \, ,
\end{equation}
acting on the space of BPS vortex fluctuations. This operator ${\cal D}$ allows us to embedd ${\cal H}^+$ in a SUSY Quantum Mechanical system because ${\cal H}^+={\cal D}^\dagger {\cal D}$ and we find the following SUSY algebra generated by the supercharge  $Q=\left(\begin{array}{cc} 0 & 0 \\ {\cal D} & 0 \end{array}\right)$:
\begin{equation}
Q^2= \left(Q^\dagger\right)^2=0 \hspace{0.5cm}, \hspace{0.5cm} {\cal H}=Q Q^\dagger+Q^\dagger Q=\left(\begin{array}{cc} {\cal H}^+={\cal D}^\dagger \, {\cal D} & 0 \\ 0 & {\cal H}^- ={\cal D} \, {\cal D}^\dagger \end{array}\right) \, , \label{susy01}
\end{equation}
${\cal H}$ being the SUSY Hamiltonian, while the SUSY partner to ${\cal H}^+$
is:
\begin{equation}
{\cal H}^- =  \left( \begin{array}{cccc}
-\Delta + |\psi|^2 & 0 & 0 & 0 \\
0 & -\Delta +|\psi|^2 & 0 & 0 \\
0 & 0 & -\Delta +\frac{1}{2} (|\psi|^2+1)+V_kV_k & -2 V_k \partial_k -\partial_k V_k \\
0 & 0 & 2V_k \partial_k + \partial_k V_k & -\Delta +\frac{1}{2} (|\psi|^2+1) + V_kV_k
 \end{array} \right) \, \, . \label{hmenos}
\end{equation}
Except for the eigenfunctions in the kernel of ${\cal D}$, which are zero modes of ${\cal H}^+$, the two operators are isospectral. This supersymmetric structure led to the proof of the Weinberg index theorem on the plane \cite{Weinberg}:
\[
{\rm ind}\,{\cal D} ={\rm dim} \, {\rm Ker} \, {\cal D}=\lim_{M\to \infty} {\rm Tr}_{L^2}\Big\{\frac{M^2}{{\cal D}^\dagger{\cal D}+M^2}-\frac{M^2}{{\cal D}{\cal D}^\dagger+M^2}\Big\}=2 n \, \, ,
\]
stating that ${\cal H}^+$ has $2n$ zero modes in its spectrum: ${\cal H}^+\xi_{0l}^+(\vec{x})=0, \, \, l=1,2, \cdots, 2n$. Moreover,  (\ref{susy01}) also guarantees that the ${\cal H}^+$-spectrum is non-negative, such that the zero modes of ${\cal D}$ are all the ground states of ${\cal H}$ because ${\rm Ker}{\cal H}^-=0$.

We shall focus on BPS cylindrically symmetric vortex filaments shaped according to the Nielsen-Olesen ansatz:
\begin{equation}
\phi(\vec{x})=f_n(r) \, e^{in\theta} \hspace{0.5cm};\hspace{0.5cm} r A_\theta(r,\theta)= n \, \beta_n(r) \label{cyan2} \, \, .
\end{equation}
We stress that: (1) Cylindrical coordinates are chosen in the $\mathbb{R}^3$-space and the vector potential components are adapted to them. (2) Besides the temporal and axial gauges, the radial gauge $A_r=0$ is assumed such that the vector field is purely vorticial. (3) The complex scalar field is expressed in polar form.

The first-order PDE system (\ref{ode11}) becomes the following ODE system:
\begin{equation}
\frac{df_n}{d r}(r)=\frac{n}{r} f_n(r) [1-\beta_n(r)]\hspace{0.5cm},\hspace{0.5cm} \frac{d\beta_n}{d r}(r)=\frac{r}{2n}[1-f_n^2(r)] \, . \label{ode1}
\end{equation}
The solutions for the radial profiles $f_n(r)$ and $\beta_n(r)$, in the $x_1:x_2$-plane and infinitely repeated along the $x_3$-axis, determine the cylindrically symmetric BPS vortex solutions. The finiteness of the energy per unit length demands that $f_n(r)\rightarrow 1$ and $\beta_n(r)\rightarrow 1$ as $r\rightarrow \infty$.

Some analytical progress in the investigation of the zero-mode fluctuations on BPS cylindrically symmetric vortices was achieved in \cite{Weinberg}. Further comprehension of their structure was obtained in References \cite{Ruback} and \cite{Burzlaff}. Here, the motivation leading several researchers to describe in detail the vortex zero modes came from the study of vortex scattering at low energies within the approach of geodesic dynamics in their moduli space, see e.g. \cite{guilarte}. In this note we shall focus on finding and describing excited fluctuation modes in the discrete ${\cal H}^+$-spectrum, i.e., internal modes of fluctuation, where the BPS vortex captures scalar and/or vector mesons, an issue not discussed in the literature on the Abelian Higgs model.

\section{Spectrum of cylindrically symmetric BPS vortex fluctuations}

In the search for positive bound states  $\xi_\lambda^+(\vec{x})$ in the discrete spectrum of the operator ${\cal H}^+$, the use of supersymmetry  is convenient. If $\omega^2_\lambda >0$, the SUSY structure (\ref{susy01}) implies that ${\cal H}^\pm \xi_\lambda^\pm (\vec{x})=\omega_\lambda^2 \xi^\pm(\vec{x})$. Moreover, the eigenfunctions of ${\cal H}^\pm$ are related through the supercharges: $\xi^+_\lambda(\vec{x})= \frac{1}{\omega_\lambda}\,{\cal D}^\dagger \xi^-_{\lambda}(\vec{x})$. In addition, it should be recalled that
the background gauge condition must be satisfied in order to eliminate spurious gauge fluctuations. The strategy is thus
to solve the spectral problem for ${\cal H}^-$ first and apply the ${\cal D}^\dagger$ operator to the $\xi^-_\lambda$ eigenfunctions, finally obtaining the eigenfunctions $\xi^+_\lambda$ of ${\cal H}^+$. This indirect path is more appropriate because the spectral problem of ${\cal H}^-$ is more tractable owing to its block-diagonal form. Two classes of eigenmodes of the operator ${\cal H}^-$ can be distinguished:

\vspace{0.2cm}

\noindent $\bullet$ \underline{Class A ${\cal H}^-$-eigenmodes}: The two $1\times 1$ block-diagonal sub-matrix differential operators in ${\cal H}^-$ prompt a complete decoupling of the vector field from the scalar fluctuations in the ${\cal H}^-$-spectral problem. There thus exist eigenfunctions of the form $[\xi_\lambda^{\rm A -}(\vec{x})]^t=\left( \begin{array}{cccc} a_1(\vec{x}) & 0 & 0 & 0 \end{array} \right)^t$ and $[\zeta_\lambda^{\rm A -}(\vec{x})]^t=\left( \begin{array}{cccc} 0 & a_2(\vec{x}) & 0 & 0 \end{array} \right)^t$. The ${\cal H}^+$-wave functions of the form ${\cal D}^\dagger [\zeta_\lambda^{\rm A -}(\vec{x})]$ do not satisfy the background gauge. Therefore, we shall study only the physically meaningful possibilities  ${\cal D}^\dagger [\xi_\lambda^{\rm A -}(\vec{x})]$ among this class of eigenmodes. The non-null component of $\xi_\lambda^{\rm A -}(\vec{x})$ complies with the PDE $(-\nabla^2 + |\psi|^2)\, a_1(\vec{x})= \omega_\lambda^2 \, a_1(\vec{x})$, or, in polar coordinates,
\begin{equation}
-\frac{\partial^2 a_1}{\partial r^2} - \frac{1}{r} \frac{\partial a_1}{\partial r} - \frac{1}{r^2} \frac{\partial^2 a_1}{\partial \theta^2} + [f_n^2(r)-\omega_\lambda^2 ] a_1=0 \label{pdeApolar} \, \, .
\end{equation}
The separation ansatz $a_1(\vec{x})=v_{nk}(r) \cos (k\theta)$ {\footnote{ $a_1(\vec{x})=v_{nk}(r) \sin (k\theta)$, $k=1,2\dots$, also leads to (\ref{ode55}). We shall pursue the cosine alternative, for the sake of brevity.} leads to the 1D Sturm-Liouville problem
\begin{equation}
-\frac{d^2 v_{nk}(r)}{d r^2} -\frac{1}{r} \frac{d v_{nk}(r)}{d r} + \Big[ f_n^2(r)-\omega_\lambda^2 + \frac{k^2}{r^2} \Big] v_{nk}(r)=0 \, \,
\label{ode55}
\end{equation}
for the radial form factor $v_{nk}(r)$. Univaluedness of the fluctuations demand that the wave number $k$ must be a natural number: $k=0,1,2\dots$. The ODE (\ref{ode55}) is no more than a radial Schr$\ddot{\rm o}$dinger differential equation with a potential well $V_{\rm eff}^{\rm A}(r;n,k)=f_n^2(r) + \frac{k^2}{r^2}$, which includes a centrifugal barrier when $k\neq 0$, bounded below and running to $1$ at infinity: $\lim_{r\rightarrow \infty} V_{\rm eff}(r)=1$. Consequently, a continuous spectrum arises in the  $\omega_\lambda^2\in [1,\infty)$ range, i.e., for energies above the scattering threshold $\omega^2_\lambda=1$. Below this threshold, in the $\omega_\lambda^2 \in (0,1)$  range, boson-vortex bound states may exist if the spectral problem (\ref{ode55}) admits eigenvalues. The procedure to find both the eigenvalues $\omega^2_j$ and the eigenfunctions $v_{nk;j}(r)$ will be implemented in the next Section for low values of $n$. The need to identify the eigenvalues will leads us to convert the ODE (\ref{ode55}), where $\omega^2_\lambda$ is a priori unknown, in a system of equations of finite differences by some discretization method of the half-line to a lattice with a finite but large number of points. Diagonalization of the matrix of the linear system in turn provides the eigenvalues and eigenfunctions, which are very good approximations to the eigenfunctions and eigenvalues of ${\cal H}^-$ provided that the number of points of the discretization is large enough.

\vspace{0.2cm}

\noindent $\bullet$ \underline{Class B ${\cal H}^-$-eigenmodes}: The $2\times 2$ block-diagonal sub-matrix in ${\cal H}^-$ acts only on scalar field fluctuations of the form $[\xi_\lambda^{\rm B-}(\vec{x})]^t=\left( \begin{array}{cccc} 0 & 0 & \varphi_1(\vec{x}) & \varphi_2(\vec{x}) \end{array} \right)^t$, leading to the spectral  PDE system:
\begin{eqnarray}
\left[-\nabla^2 + {\textstyle \frac{1}{2}} \, ( |\psi|^2+1) + V_k V_k \right] \varphi_1(\vec{x}) - 2 \,V_k \,\partial_k \varphi_2 (\vec{x})&=& \omega_\lambda^2 \, \varphi_1(\vec{x}) \label{odeC1}\\
\left[-\nabla^2 + {\textstyle \frac{1}{2}} \, ( |\psi|^2+1) + V_k V_k \right] \varphi_2(\vec{x}) + 2\, V_k \,\partial_k \varphi_1 (\vec{x})&=& \omega_\lambda^2 \, \varphi_2(\vec{x}) \label{odeC2} \, \, .
\end{eqnarray}
For cylindrically symmetric BPS vortices, the ansatz $\varphi_1(\vec{x}) = - \, u_{nk}(r) \sin[(k+1)\theta]$, $\varphi_2(\vec{x})=u_{nk}(r) \cos[(k+1)\theta]$ that converts the PDE system (\ref{odeC1}) and (\ref{odeC2}) into the spectral ODE is clear:
\begin{equation}
{\cal R}u_{nk}(r) = -\frac{d^2 u_{nk}(r)}{d r^2} - \frac{1}{r} \frac{d u_{nk}}{d r} + \Big[ \frac{1}{2} (1+f_n^2(r)) + \frac{(n \beta_n(r)-(1+k))^2}{r^2} -\omega_\lambda^2 \Big]u_{nk}(r) =0 \label{odeC3} \quad ,
\end{equation}
where the radial form factor $u_{nk}(r)$ is the unknown. This is a radial Schr$\ddot{\rm o}$dinger differential equation with an effective potential well: $V_{\rm eff}^{\rm B}(r)=\frac{1}{2} [1+f_n^2(r)] + \frac{1}{r^2}[n \beta_n(r)-(1+k)]^2$. From the functional behavior of $V^{\rm B}_{\rm eff}$, we may conclude that a continuous spectrum arises settled on the threshold value $\omega_\lambda^2=1$. The same reasoning indicates that there could exist bound states in this class with eigenvalues in the $\omega\in (0,1)$ range. However, a theoretical argument can be used to discard this possibility. The linear differential operator ${\cal R}$ associated with the ODE (\ref{odeC3}) can be factorized as ${\cal R}-1= {\cal L}^\dagger {\cal L}$, where ${\cal L}$ refers to the first-order differential operators ${\cal L}=-\frac{d}{d r} +
\frac{1}{r}[1+k-n\beta_n(r)]$. This means that the ${\cal R}-1$ operator has a non-negative spectrum and, consequently, there are no bound states in the discrete spectrum of ${\cal H}^-$ within this class B of fluctuations.

\vspace{0.2cm}

Translation via use of the supercharges of all the ${\cal H}^-$-spectral information described  previously reveals the structure of the ${\cal H}^+$-spectrum:

\vspace{0.2cm}

\noindent $\bullet$ \underline{Class A ${\cal H}^+$-eigenmodes}: Assuming knowledge of $\omega_\lambda^2$ and $v_{nk}(r)$ from the solution of (\ref{ode55}) the eigenfunctions of ${\cal H}^+$ paired through supersymmetry with these class A ${\cal H}^-$-eigenmodes take the form:
\begin{equation}
\xi^{{\rm A}+}_\lambda(\vec{x},n,k)= \left( \begin{array}{c} \sin \theta \cos (k\theta) \frac{\partial v_{nk}(r)}{\partial r} - \frac{k}{r} \, v_{nk}(r) \cos \theta \sin(k\theta) \\ -\cos \theta \cos (k\theta) \frac{\partial v_{nk}(r)}{\partial r} - \frac{k}{r} \, v_{nk}(r) \, \sin \theta \sin(k\theta) \\ f_n(r) \, v_{nk}(r)\, \cos(n\theta)\, \cos(k\theta) \\ f_n(r)\, v_{nk}(r)\, \sin(n\theta) \,\cos(k\theta)
 \end{array} \right) \hspace{0.2cm} , \hspace{0.2cm} k=0,1,2,\dots\, . \hspace{0.2cm} \label{excitedmode1}
\end{equation}
It is immediate to check that $\xi_\lambda^{\rm A+}(\vec{x})$ satisfies the background gauge (\ref{backgroundgauge}), meaning that the fluctuations (\ref{excitedmode1}) correspond to admissible eigenfunctions of the second-order BPS vortex fluctuation operator ${\cal H}^+$ \footnote{The sine alternative leads to degenerate eigenfunctions with $\xi_\lambda^{A+}(\vec{x},n,k)$ obtained by simply replacing  $\cos(k\theta)$ by $\sin(k\theta)$ and $\sin(k\theta)$ by $-\cos(k\theta)$ in the wave function (\ref{excitedmode1}). Knowledge of the radial form factor $v_{nk}(r)$ authomatically gives both the ${\rm cos}$ and ${\rm sin}$ bound states.}. From (\ref{ode55}) we may conclude that a continuous spectrum emerges at the threshold value $\omega^2=1$, while the discrete spectrum is confined to the open interval $\omega^2_j\in (0,1)$. The presence of bound states in the ${\cal H}^+$-spectrum, however, requires the existence of eigenfunctions of (\ref{ode55}) satisfying the boundary conditions $\frac{d v_{nk}}{dr}(0)=0$ and $\lim_{r\rightarrow \infty} v_{nk}(r)=0$. This point will be addressed in the next section.

\vspace{0.2cm}

\noindent $\bullet$ \underline{Class B ${\cal H}^+$-eigenmodes}: The corresponding SUSY partner ${\cal H}^+$-eigenfunctions $\xi_\lambda^{\rm B+}(\vec{x})$ associated with the class B ${\cal H}^-$-eigenfunctions are given by:
\[
\xi_\lambda^{\rm B+}(\vec{x})= r^{n-k-1} \left( \begin{array}{c}
h_{nk}(r) \sin [(n-k-1)\theta] \\
h_{nk}(r) \cos[(n-k-1)\theta] \\
-\frac{h_{nk}'(r)}{f_n(r)} \cos (k\theta) \\
-\frac{h_{nk}'(r)}{f_n(r)} \sin (k\theta)
\end{array} \right) \hspace{0.5cm} .
\]
The new radial form factor $h_{nk}(r)$ may be defined from the relation  $u_{nk}(r)=\frac{r^{n-k-1}}{f_n(r)} h_{nk}(r)$ in such a way that the ODE (\ref{odeC3}) turns into the equation
\[
r\,h_{nk}''(r) + [-1-2k+2n \beta_n(r)] h_{nk}'(r) +r [\omega_\lambda^2 -f_n^2(r)]h_{nk}(r)=0 \, \, ,
\]
for the radial form factor $h_{nk}(r)$. Regularity at the origin, however, of the positive eigenfluctuations $\xi_\lambda^{\rm B+}(\vec{x})$ requires that $0\leq k \leq n-1$. In this case, there exists a continuous spectrum emerging from the threshold value $\omega^2=1$ and no positive eigenvalue bound states arise. Notice, however, that the form of these eigenmodes $\xi_\lambda^{\rm B+}(\vec{x})$ follow the ansatz given in \cite{Ruback} for the zero modes, such that we can regard the $2n$ zero modes as the only bound class B ${\cal H}^+$-eigenmodes {\footnote{Other $n$ zero modes are easily generated by rotating $\pi/2$ separately in the scalar and vector field fluctuations, which is a symmetry of the spectral problem. The same argument can be applied to the eigenfunctions $\xi_\lambda^{B+}(\vec{x})$.}}.

Orthogonality between eigenfunctions belonging to different classes is guaranteed by the conservation of the scalar product in the SUSY partnership: $\left\langle \xi_\lambda^{A+}(\vec{x}) , \xi_{\lambda'}^{B+}(\vec{x}) \right\rangle = \left\langle \xi_\lambda^{A-}(\vec{x}) , \xi_{\lambda'}^{B-}(\vec{x}) \right\rangle =0$, because class A and B eigenfunctions of ${\cal H}^-$ are clearly orthogonal.  Orthogonality between eigenfunctions belonging to the same class with different angular dependence is established by Fourier analysis.

\section{Positive eigenvalue bound states of the small vortex fluctuation operator}

We now attempt to elucidate the existence of excited fluctuations of class A belonging to the discrete spectrum of ${\cal H}^+$ with positive eigenvalues lower than $1$. The search for and the analysis of these fluctuations reduce to the numerical computation of the radial form factor $v_{nk}(r)$ in the ODE (\ref{ode55}). Our strategy to achieve this is to employ a second-order finite-difference scheme that simulates the differential equation (\ref{ode55}) by the recurrence relations
\begin{equation}
-\frac{v_{nk;j}^{(i+1)} - 2 v_{nk;j}^{(i)} +v_{nk;j}^{(i-1)}}{(\Delta x)^2} - \frac{v_{nk;j}^{(i+1)} - v_{nk;j}^{(i-1)}}{2i (\Delta x)^2} + \Big[ f_n^2(i \Delta x) + \frac{k^2}{i^2 (\Delta x)^2} \Big] v_{nk;j}^{(i)} = \omega_{nk;j}^2 \, v_{nk;j}^{(i)} \label{erec1} \quad ,
\end{equation}
where we have confined the problem to the interval $[0,r_{\rm max}]$ for a large enough $r_{\rm max}$. We denote $v_{nk;j}^{(i)}=v_{nk;j}(i\Delta x)$, with $\Delta x= \frac{r_{\rm max}}{N}$, and choose a mesh of $N$ points with $i=0,1,2, \cdots , N$. The eigenfunctions and the eigenvalues depend on the values of the angular momentum $k$ and the vorticity $n$. The index $j$ is used to enumerate the discrete eigenfunctions. The contour conditions are:
\[
\textbf{(1)} \hspace{0.5cm} -\frac{4}{3} \frac{v_{nk;j}^{(2)}-v_{nk;j}^{(1)}}{(\Delta x)^2} + \Big[f_n^2 (\Delta x) + \frac{k^2}{(\Delta x)^2} \Big] v_{nk;j}^{(1)} = \omega_{nk;j}^2 \, v_{nk;j}^{(1)}\, \, \hspace{0.5cm} \mbox{and} \hspace{0.5cm} \textbf{(2)} \hspace{0.5cm} v_{nk;j}^{(N)}=0
\]
A good estimation of the discrete eigenvalues $\omega_{nk;j}^2$ is obtained through diagonalization of the $N\times N$ matrix in the left member of the linear system (\ref{erec1}). We show the eigenvalues of ${\cal H}^+$ for low values of $n$ and $k$ obtained in a Mathematica environment in Table 1 by applying this procedure with the choice of $N=400$. In Figure 1 we illustrate the behavior of the potential wells of the radial Schr\"odinger equation (\ref{ode55}) for $n=4$ and the overlapped dashed lines that determine the discrete eigenvalues of the ${\cal H}^+$-spectrum in this case. The radial form factor $v_{nk;j}(r)$ of the eigenfunctions associated with these eigenvalues are also shown.

In general, we observe that the number of bound states increases with the magnetic flux $n$. In particular, we conclude the existence of 1 bound state for $n=1$-vortices; 2 bound states for the $n=2$ and $n=3$ vortices; 3 bound states for $n=4$-vortices, and 4 bound states in the case of $n=5$-vortices. In Table 1 we include a graphical representation of the discrete ${\cal H}^+$-spectrum for several values of the vorticity $n$.

\begin{table}[h]
\begin{tabular}{cc}
\begin{tabular}{|c||c|c|c|} \hline
\multicolumn{4}{|c|}{\rule[-0.3cm]{0.0cm}{0.8cm}Eigenvalues of the discrete spectrum of ${\cal H}^+$} \\ \hline\hline
$n$ & \rule[-0.3cm]{0.0cm}{0.8cm} $k=0$ & $k=1$ & $k=2$ \\ \hline\hline
$1$ &  \rule[-0.3cm]{0.0cm}{0.8cm} $(\omega_{10;1}^{\rm A})^2=0.777446$ & - & -  \\ \hline
$2$ & \rule[-0.3cm]{0.0cm}{0.8cm}  $(\omega_{20;1}^{\rm A})^2=0.538573$ & $(\omega_{21;1}^{\rm A})^2=0.972563$ & - \\ \hline
$3$ & \rule[-0.3cm]{0.0cm}{0.8cm}  $(\omega_{30;1}^{\rm A})^2=0.402692$ & $(\omega_{31;1}^{\rm A})^2=0.830078$ & - \\ \hline
$4$ & \rule[-0.5cm]{0.0cm}{1.2cm} $\begin{array}{c} (\omega_{40;1}^{\rm A})^2=0.319276 \\ (\omega_{40;2}^{\rm A})^2=0.988212 \end{array}$ & $(\omega_{41;1}^{\rm A})^2=0.701708$ & -  \\ \hline
$5$ & \rule[-0.5cm]{0.0cm}{1.2cm} $\begin{array}{c} (\omega_{50;1}^{\rm A})^2=0.263671 \\ (\omega_{50;2}^{\rm A})^2=0.939461 \end{array}$ & $(\omega_{51;1}^{\rm A})^2=0.601223$ & $(\omega_{52;1}^{\rm A})^2 = 0.942438$  \\ \hline
\end{tabular} & \begin{tabular}{c}
\includegraphics[width=4cm]{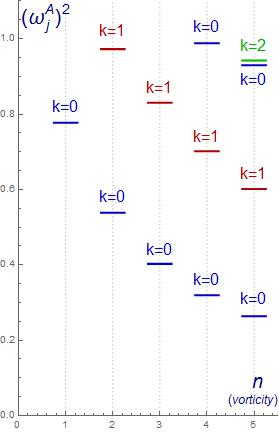}
\end{tabular}
\end{tabular}

\caption{Numerical estimation of the discrete spectrum eigenvalues associated with the class A eigenfunctions $\xi_\lambda^{\rm A+}(\vec{x},n,k)$ with angular momentum $k$, together with a graphical representation of the second-order small $n$-vortex fluctuation operator ${\cal H}^+$ spectrum. The eigenvalues of the form $\omega_{n0}$ shown in this Table agree with those given in Table 3 of Reference \cite{Hindmarsh} for critical quotient of the scalar and vector field masses up to a factor of $2$ due to a different convention.}
\end{table}

\begin{figure}[ht]
\centering
\includegraphics[height=3.5cm]{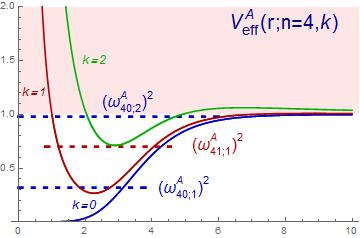} \hspace{2cm}  \includegraphics[height=3.5cm]{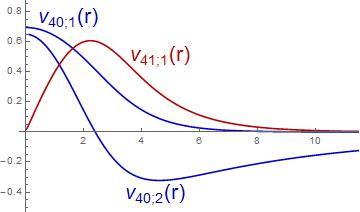}

\caption{Plots of the effective radial potential wells $V_{\rm eff}^{\rm A}(r,n=4,k)$ arising in (\ref{ode55}) for $k=0,1,2$ (solid lines) with the discrete eigenvalues (dashed lines) overlapped (left) and the corresponding radial eigenfunctions $v_{4k,j}(r)$ (right) of the small $n=4$ vortex fluctuation operator ${\cal H}^+$.}
\end{figure}

In sum, there exist bound states $\xi_{nk,j}^{\rm A+}(\vec{x})$ that are eigenfunctions of ${\cal H}^+$. Here, we have described
the stationary wave functions where an awkward combination of scalar and vector boson fluctuations are trapped by a cylindrically symmetric BPS vortex. These configurations oscillate in time with frequencies determined by the discrete eigenvalues and are thus internal modes of fluctuation of the BPS $n$-vortex.

\end{document}